\documentclass[prb,twocolumn,superscriptaddress]{revtex4-1}
\usepackage{amsmath,amssymb,amsfonts,bm,graphicx}
\usepackage{braket}

\usepackage{graphicx}
\usepackage{amsmath}
\usepackage{bm}

\usepackage[T1]{fontenc}
\usepackage{fourier}
\usepackage{baskervald}

\graphicspath{{figs/}}

\begin{document}

\title{Spatial inhomogeneity and the metal-insulator transition 
in Ca$_3$(Ru$_{1-x}$Ti$_x$)$_2$O$_7$}

\author{Frank Lechermann}
\affiliation{I. Institut f\"ur Theoretische Physik, Universit\"at Hamburg, 
Jungiusstrasse 9, 20355 Hamburg, Germany}

\author{Qiang Han}
\affiliation{Department of Physics, Columbia University,
	538 West 120th Street, New York, NY 10027, USA}

\author{Andrew J. Millis}
\affiliation{Department of Physics, Columbia University,
	538 West 120th Street, New York, NY 10027, USA}
\affiliation{Center for Computational Quantum Physics, 
The Flatiron Institute, 162 5th Avenue, New York, NY 10010, USA}
\date{\today}

\begin{abstract}
Turning a pristine Mott insulator into a correlated metal by chemical doping
is a common procedure in strongly correlated materials physics, e.g. underlying
the phenomenology of high-$T_c$ cuprates. The ruthenate bilayer compound 
Ca$_3$Ru$_2$O$_7$ is a prominent example of a reversed case, namely a correlated 
metal at stoichiometry that realizes a transition into an insulating state via 
Ti doping. We here investigate this puzzling metal-insulator transition (MIT) by 
first-principles many-body theory and elucidate a challenging interplay between 
electronic correlations and symmetry breakings on the Ru sublattice. While average 
effects on the Ca$_3$Ru$_2$O$_7$ crystal structure are still relevant, key to the MIT 
is the cooperation of electronic correlations with the spatial inhomogeneity in the 
defect regime. Together they give rise to the emergence of site-selective Mott 
criticality and competing orbital-ordering tendencies. 
\end{abstract}

\maketitle

{\it Introduction.---}
The interplay of various degrees of freedom, e.g. of charge, orbital, spin or structural
kind, is key to an understanding of many realistic metal-insulator transitions (MITs)
in nature~\cite{han18}. 
In this respect, the Ruddlesden-Popper series of $n$-layered calcium ruthenates 
Ca$_{n+1}$Ru$_n$O$_{3n+1}$ poses a particularly challenging problem. It is agreed that
electronic correlations arise from low-spin Ru$^{4+}(4d^4)$ sites in these 
compounds, formally locating them in a Hund-metal~\cite{wer08,hau09,med11} regime. 
While the $n\rightarrow\infty$ perovskite CaRuO$_3$ is metallic~\cite{han16} with 
competing magnetic interactions~\cite{he01} down to lowest temperatures, distorted 
Ca$_2$RuO$_4$ ($n$=1) undergoes a paramagnetic MIT at 
$T_{\rm MIT}=357$\,K and displays antiferromagnetic (AFM) order below 
$T_{\rm N}=110$\,K~\cite{ale99}. From these limiting cases, Mott criticality is 
expectedly intricate in the bilayer system. And indeed, though Ca$_3$Ru$_2$O$_7$ shows 
several electronic and magnetic transitions for 
$T<100$\,K~\cite{cao97,yos04,bau06,lee07,pen13}, a robust insulating state is not 
reached at stoichiometry. With its non-centrosymmetric $Bb2_1m$ space group, the 
bilayer ruthenate marks the case of a 'polar metal'~\cite{yos05,sin06,tho18}, and has 
gained recent strong interest due to a complex fermiology~\cite{bau06,hor19,mar20,pug20}.

Doping of about 5\% titanium gives rise to a MIT in the bilayer 
compound~\cite{ke11,pen13} at $T_{\rm MIT}\sim 80$\,K. The magnetic order 
switches from the stoichiometric A-type ordering of AFM-coupled ferromagnetic 
bilayers to G-type AFM ordering. There is apparently no paramagnetic Mott-insulating 
phase in Ca$_3$(Ru$_{1-x}$Ti$_x$)$_2$O$_7$. Weakly-localized states are already observed 
for very small Ti doping~\cite{tsu13}, and perculative behavior is also 
detected~\cite{pen13}. 
According to Ke {\sl et al.}~\cite{ke11}, the substitutional dopants enter as 
Ti$^{4+}(3d^0)$ impurities, therefore do not provide any significant 
charge doping. Hence the doping-induced MIT has originally been associated 
with the blocking of hopping paths~\cite{ke11}. Furthermore, the averaged 
crystal structure of the Ti-doped bilayer~\cite{pen13} displays also an enhanced 
two-dimensionality of the bilayers as well as a somewhat increased tilting of the RuO$_6$ 
octahedra. These global structural changes lead to a larger averaged crystal-field (CF) 
splitting $\Delta$ between the three partially occupied Ru$(4d)$ states $m=xy,xz/yz$ of 
$t_{2g}$ character. In fact, a large $\Delta$ is the major driving force for the MIT in 
Ca$_2$RuO$_4$~\cite{gor10,lie07}. Since the energy scales in Ca$_3$Ru$_2$O$_7$ are generally 
smaller than in the latter single-layer ruthenate, the competition between the various 
degrees of freedom is much more subtle.

In this work, the goal is to unveil the detailed cooperation of defect physics and 
electronic correlations that drive the MIT in Ti-doped Ca$_3$Ru$_2$O$_7$. By means of 
the real-space combination of density functional theory (DFT) and dynamical mean-field
theory (DMFT) applied to a defect supercell with 6.25\% Ti concentration in a
fully charge self-consistent manner, we profoundly account for average and local effects
on an equal footing. While average effects from doping are notable, the spatial 
inhomogeneity introduced by Ti defects and cooperating with electron correlations
is the crucial driving force towards the insulating phase.

{\it Theoretical approach.---}
Charge self-consistent DFT+DMFT~\cite{sav01,pou07,gri12} is used to access the 
correlated electronic structure. For the DFT part, a mixed-basis pseudopotential 
method~\cite{lec02,mbpp_code}, employing the generalized-gradient 
approximation in the Perdew-Burke-Ernzerhof form~\cite{per96}, is put into practise.
Spin-orbit coupling is neglected in this work. The multi-single-site DMFT impurity 
problems encountered in the basic unit cell as well as in the defect supercell are 
solved by the hybridization-expansion continuous-time quantum Monte Carlo 
scheme~\cite{wer06}, as implemented in the TRIQS code~\cite{par15,set16}. The 
correlated subspace consists of the effective transition-metal (TM) $t_{2g}$ 
Wannier-like functions $w_m(t_{2g})$, i.e. covers locally three orbitals. These 
functions are obtained from the projected-local-orbital formalism~\cite{ama08,ani05}, 
using as projection functions the linear combinations of atomic-like $t_{2g}$ orbitals 
that diagonalize the TM local $w_m(t_{2g})$-orbital density matrix on each site. 

Local Coulomb interactions in general Slater-Kanamori form, i.e. including density-density 
as well as spin-flip and pair-hopping terms, are parametrized by a Hubbard $U$ 
and a Hund's exchange $J_{\rm H}$. The intraorbital interaction $U$ on the Ru sites is 
treated as a parameter, ranging at most from 1.5\,eV to 7.5\,eV. The Hund's exchange 
is chosen $J_{\rm H}=0.4$\,eV for $U<2$\,eV and $J_{\rm H}=0.7$\,eV for $U>2$\,eV, 
in order to allow for comparison with previous theory work on ruthenates~\cite{gor10}. 
For the Ti sites, a value $U=5$\,eV (and $J_{\rm H}=0.7$\,eV) is chosen to account for 
the fact that interactions in the Ti$(3d)$ shell may be larger than in the Ru$(4d)$ shell. 
To obtain the spectral information, analytical continuation from Matsubara space by the 
maximum-entropy method and the Pad{\'e} method is performed. 
\begin{figure}[b]
\includegraphics*[width=8.5cm]{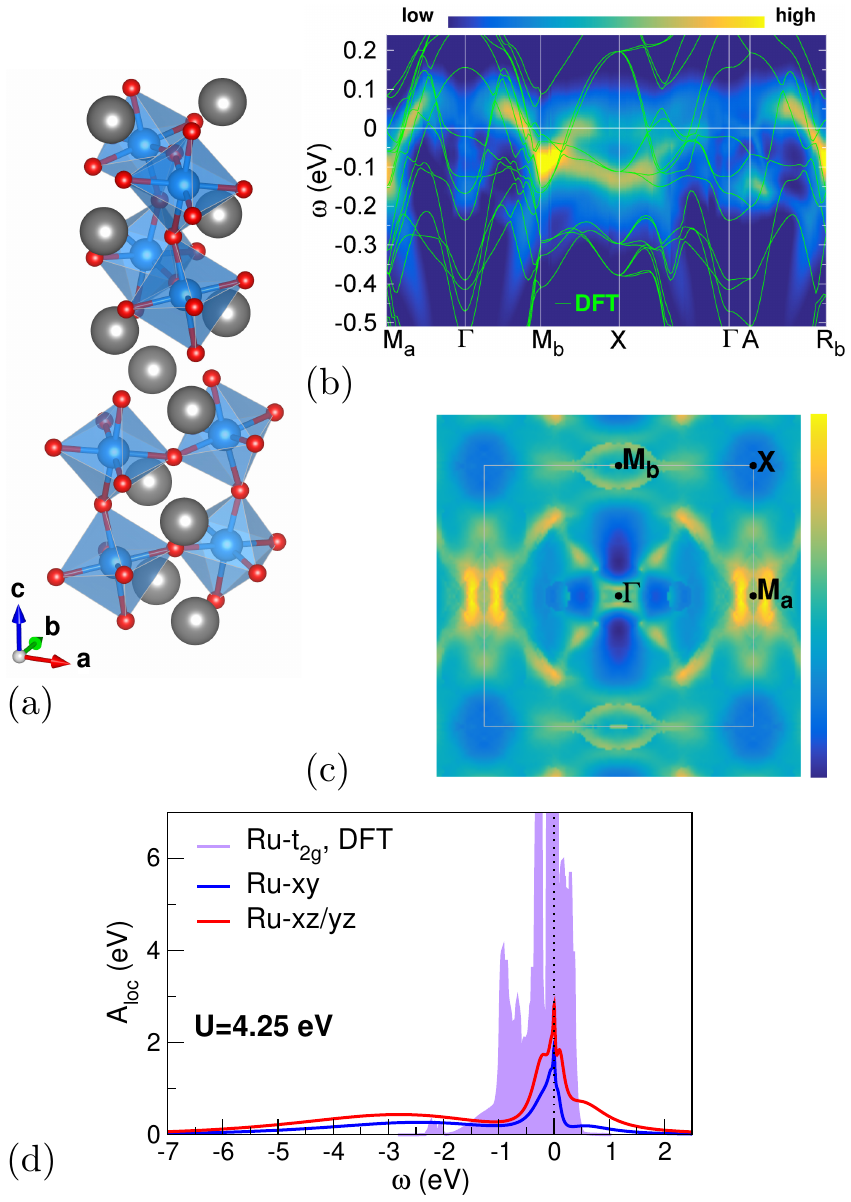}
\caption{(color online) Characterization of stoichiometric Ca$_3$Ru$_2$O$_7$. 
(a) Unit cell with two bilayers; Ca (large grey), Ru (blue) and O (small red).
(b-d) Paramagnetic DFT+DMFT results for $U=4.25$\,eV at $T=100$\,K.
(b) Spectral function $A({\bf k},\omega)$ along high-symmetry lines, 
compared to DFT bands (green). 
(c) Fermi surface in $k_z=0$ plane, gray rectangular marks the Brillouin zone.
(d) Local Ru-$t_{2g}$ spectral function discriminating the $xy$ and $xz/yz$ orbitals
and comparing to DFT spectrum.}
\label{fig:stoich}
\end{figure}

{\it Stoichiometric Ca$_3$Ru$_2$O$_7$.---}
To set the stage, we start with the correlated electronic structure of stoichiometric
Ca$_3$Ru$_2$O$_7$. Figure~\ref{fig:stoich}a displays the experimental unit 
cell~\cite{yos05} at 40\,K with inplane lattice parameters $a=5.37$\,\AA, $b=5.54$\,\AA\,
and $c=19.52$\,\AA. It incorporates two bilayers along the $c$-axis and a total of 8 Ru 
sites. The relevant CF splitting $\Delta$ in the $t_{2g}$ manifold of the Ru$(4d)$ states 
is defined as $\Delta=\varepsilon_{xy}-(\varepsilon_{xz}+\varepsilon_{yz})/2$, where 
$\varepsilon_m$ are the CF levels of the $xy,xz,yz$ Wannier orbitals. Effects due to
the small splitting between $xz$, $yz$ states are minor and will not be discussed 
henceforth (though the corresponding splitting is surely included in the calculations). 
On the DFT level, the bare CF splitting amounts to $\Delta=-153$\,meV, i.e. the $xy$ state 
is lowest in energy. This results in an orbital polarization, giving 
rise to occupations $n_{xy}=1.43$ and $n_{xz/yz}=1.26$ per orbital. Note that the given 
CF splitting is about half the value of $\Delta\sim -320$\,meV in single-layer 
Ca$_2$RuO$_4$~\cite{gor10}.

The correlated spectral data at $T=100$\,K is depicted in Figs.~\ref{fig:stoich}b-d. 
The compound is metallic, but compared to DFT,
strong renormalization and loss of coherence is revealed at low energy, in accordance with 
angle-resolved photoemission~\cite{bau06,hor19,mar20}. Significant inplane anisotropy
in the correlated fermiology takes place between M$_{\rm a}$ and M$_{\rm b}$, in
line with experiment~\cite{bau06}. A more thorough investigation of further 
lower-temperature dispersions asks for an inclusion of spin-orbit coupling, and is not
an objective of the present study. Finally, the local spectral functions show that
the $xz/yz$ states are stronger correlated with a lower Hubbard band of enhanced
weight and deeper energy location of $\sim -2.9$\,eV than the $xy$ states.
\begin{figure}[t]
\includegraphics*[width=8.5cm]{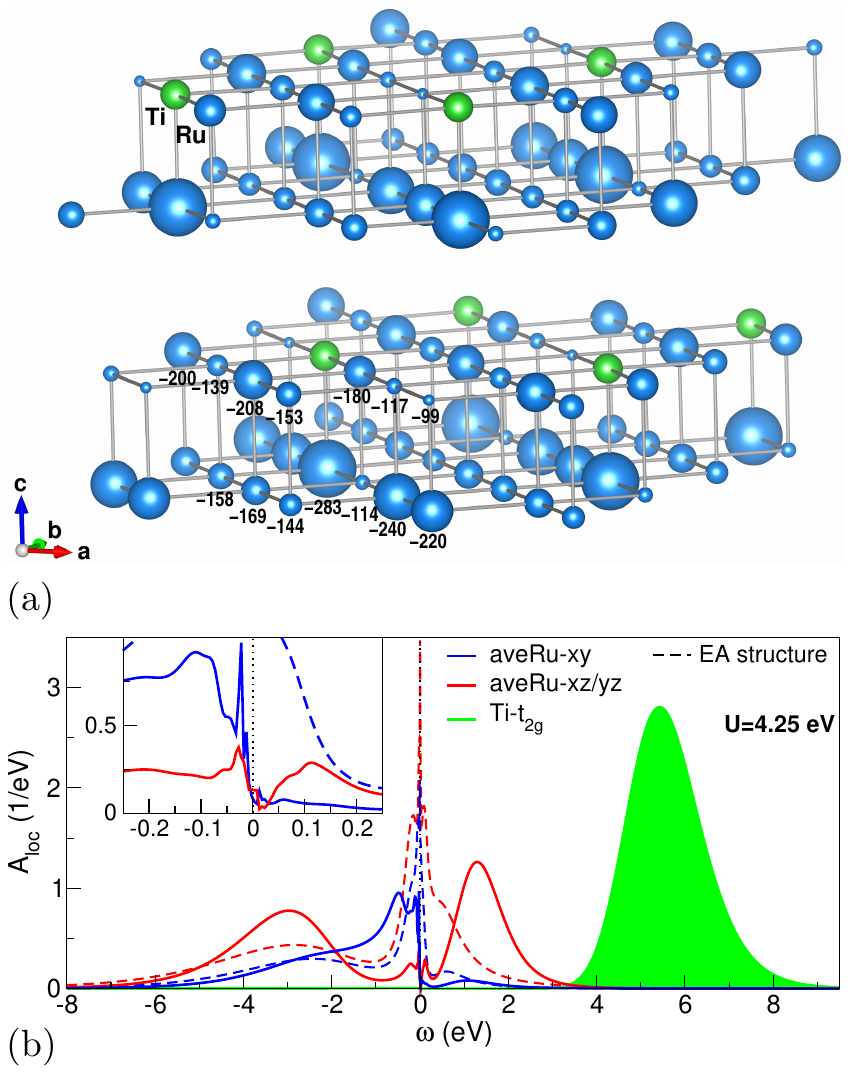}
\caption{(color online) Structure and global spectrum of 
Ca$_3$(Ru$_{0.9375}$Ti$_{0.0625}$)$_2$O$_7$.
(a) Structurally relaxed supercell showing Ru (blue) and Ti (green) ions,
indicating the crystal-field splitting $\Delta$ via the diameter of the Ru-site 
spheres (Ca and O sites are not shown). Numbers indicate the 
actual $\Delta$ value of a given Ru-site symmetry class (in meV). The diameter 
of the Ti spheres is chosen as the average $\Delta$ of the Ru sites.
(b) Average Ru$(4d)$-$t_{2g}$ and Ti$(3d)$-$t_{2g}$ paramagnetic local spectrum 
with 6.25\% Ti for $U=4.25$\,eV ($T=40$\,K). Dashed lines show the corresponding 
Ru-$t_{2g}$ spectrum of the EA structure for comparison. Inset: blow up of the
low-energy region.}
\label{fig:full}
\end{figure}

{\it Local electronic structure of Ca$_3$(Ru$_{0.9375}$Ti$_{0.0625}$)$_2$O$_7$.---}
In order to describe the bilayer ruthenate with finite Ti doping, we start from the 
experimentally-averaged (EA) structure at 10\,K with 5\% Ti doping~\cite{ke11}. The
corresponding experimental system is insulating below $T_{\rm MIT}\sim 80$\,K. As an 
averaged structure, the atom-number size of the primitive cell is identical to the one at 
stoichiometry, but with different lattice parameters $a=5.38$\,\AA, $b=5.56$\,\AA\,
and $c=19.37$\,\AA, as well as modified atomic positions. 
Thus the effect of Ti impurities is only taken into account implicitly via an averaged 
structure modification. The effective CF splitting in this structure amounts to 
$\Delta=-171$\,eV in DFT, hence is about 20\,meV larger than in the stoichiometric case. 
This is mainly attributed to an enhanced two-dimensionality and increaced tilting of the 
RuO$_6$ octahedra. 
It can be seen from the dashed lines in Fig.~\ref{fig:full}b that this treatment of Ti 
doping is not sufficient to render the system insulating for a reasonable value of 
$U=4.25$\,eV, but electronic correlations are somewhat enhanced compared to the 
stoichiometric case.
\begin{figure}[t]
\includegraphics*[width=8cm]{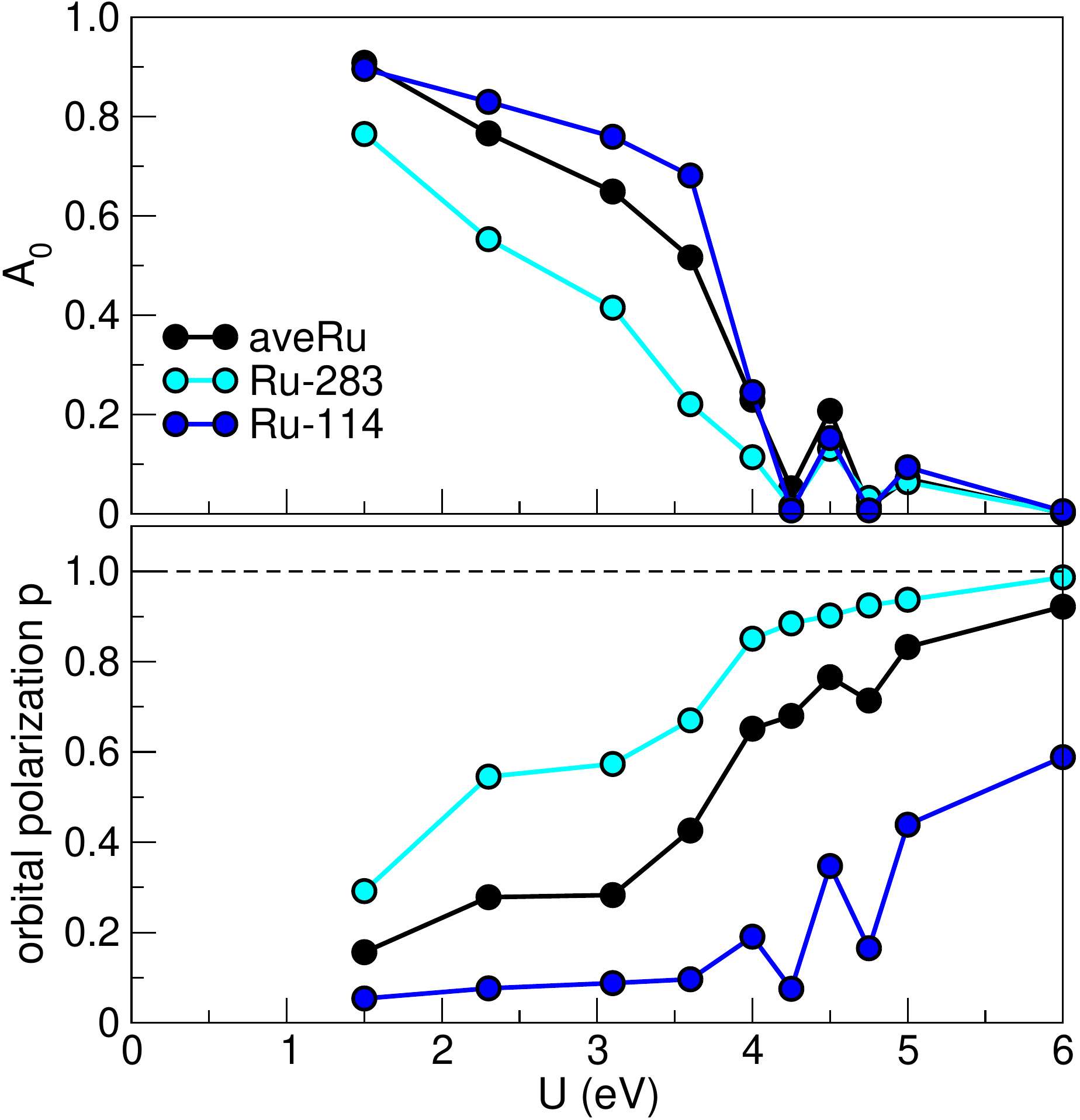}
\caption{(color online) The Ru$(4d)$-$t_{2g}$ spectral weight $A_0$ at the Fermi 
level (top) and the orbital polarization $p$ (bottom) with increasing $U$, for selected 
Ru-site classes in the Ti-doped supercell, respectively, at $T=40$\,K.}	
\label{fig:us}
\end{figure}

Let us turn to the supercell description of Ti-doped Ca$_3$Ru$_2$O$_7$. A 192-atom-site 
cell (cf. Fig.~\ref{fig:full}a) is constructed, starting from the EA structure and
introducing two Ti impurities in adjacent bilayers, i.e. each bilayer carries one
substitutional Ti defect. There are 32 TM sites in the defect supercell, 30 of Ru and 
2 of Ti kind. Fixing the scaled EA lattice parameters, we structurally relax this 
supercell within DFT+U assuming G-type AFM order. No symmetry constraints are enforced 
in the structural relaxation. This leads to site-symmetry breakings, not unexpected in 
this puzzling polar-metal system~\cite{gan20}. For the DFT+DMFT investigation at a
system temperature of $T=40$\,K, we identify 14 Ru-site classes as symmetry inequivalent. 
Together with both symmetry-equivalent Ti sites, there are hence 15 coupled impurity
problems solved at each self-consistency step. 

As in previous DFT calculations~\cite{ke11}, the Ti impurities are indeed of 
Ti$^{4+}(3d^0)$ kind, and the $t_{2g}$ electronic spectrum is accordingly located high in 
energy within the unoccupied region (see green part in Fig.~\ref{fig:full}b). Note that 
local structural relaxation shifts the Ti spectrum to somewhat smaller energies. The 
resulting CF splitting on the Ru sites is however distributed over a surprisingly large
energy window $[-99,-283]$\,eV. In order to simplify notation, we will in the following 
address the different Ru-site classes as 'Ru$\Delta$'. Interestingly, the Ru sites just 
below the Ti impurities have the largest CF splitting. These Ru-283 sites show 
a comparatively large relaxation {\sl away} from Ti. The latter may be explained by the 
fact that due to the $3d^0$ character of titanium, the effective $xz/yz$ hopping 
perpendicular to the plane is strongly weakened, resulting also in an overall reduced 
Ru-Ti bonding. Hence, the original effect of Ti sites blocking hopping paths is affirmed, 
however this single-particle-based mechanism alone cannot be sufficient to drive the MIT
at the given comparatively small amounts of doping.
\begin{figure}[t]
\includegraphics*[width=8.5cm]{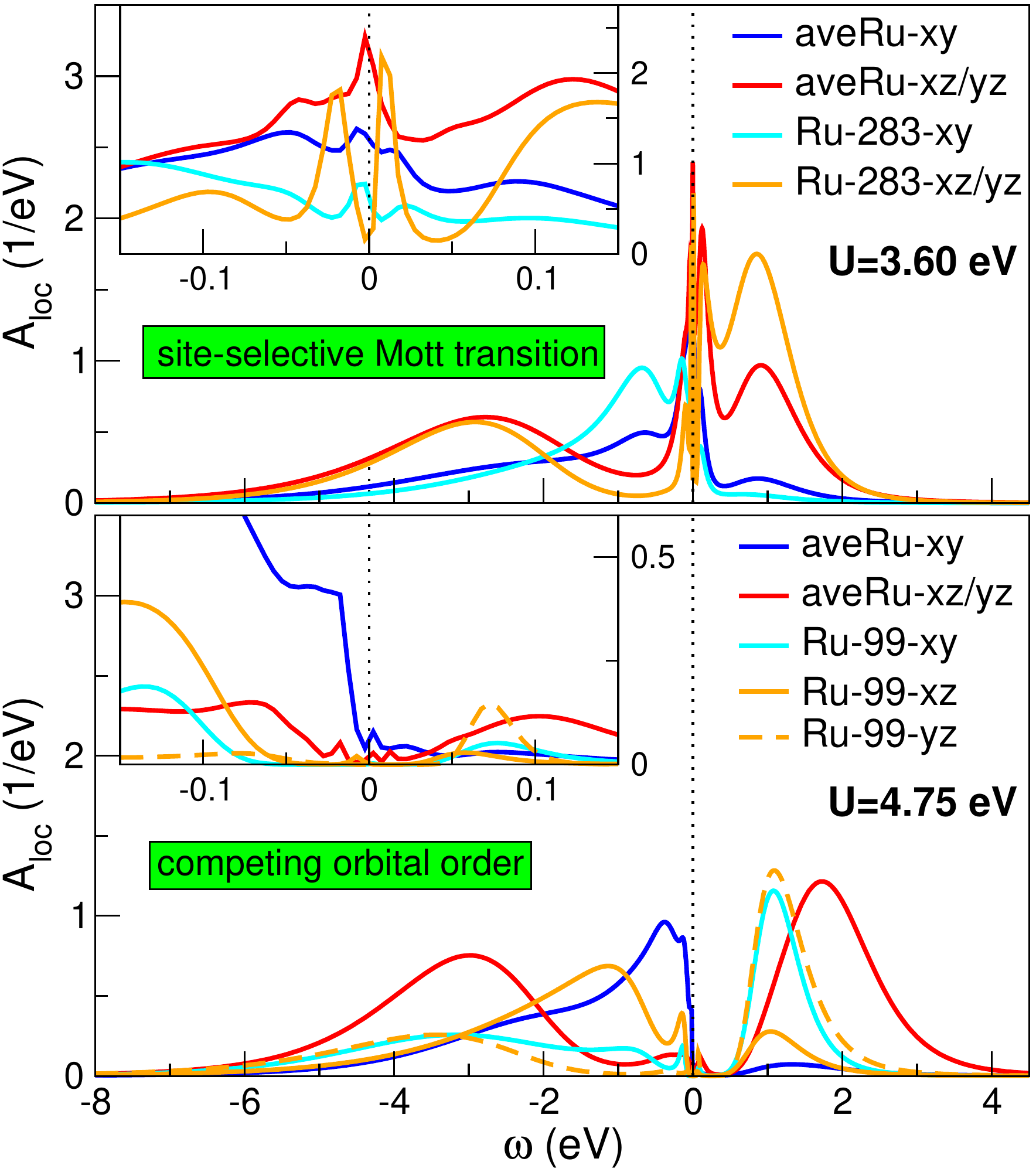}
\caption{(color online) Local Ru$(4d)$-$t_{2g}$ spectral function for $U=3.60$\,eV (top) 
and $U=4.75$\,eV (bottom) at $T=40$\,K. The average spectra and two selected Ru-site 
class spectra are shown, respectively: For Ru-283(Ru-99) at the lower(larger) 
$U$ value.}	
\label{fig:ex}
\end{figure}

Notably, the average DFT crystal-field splitting in the supercell amounts to 
$\Delta_{\rm av}=-172$\,eV and is thus indeed identical to the one in the starting EA 
structure. But for equal local Couloumb interactions, the system is Mott insulating with 
strong orbital polarization as shown in Fig.~\ref{fig:full}b. On average, the $xy$ 
state becomes fully occupied and the $xz/yz$ states each host one electron, such that 
the four-electron occupation of Ru$(4d)$-$t_{2g}$ is realized. This orbital-polarization 
scenario is reminiscent of the one in single-layer Cu$_2$RuO$_4$~\cite{gor10}. 
But since the average CF splittings of the EA structure and the aligned supercell 
correspond to each other, the spatial inhomogeneity has to play a key role in the bilayer
MIT. 

The site-selective data shown in Fig.~\ref{fig:us} and corresponding electronic
spectra in Fig.~\ref{fig:ex} render indeed obvious, that the various Ru sites behave 
quite differently, in connection with their respective $\Delta$ value. In order to
not only rely on analytical continuation, Ru local spectral weight at lowest energy, 
i.e. $A_{\rm loc}(\omega)$ at zero frequency $\omega=0$, is plotted in the top of 
Fig.~\ref{fig:us} in its approximate form 
$A_0=-\frac{\beta}{\pi}\sum_m{\rm Im}\,G^{(m)}_{\rm loc}(\beta/2)$, 
where $G^{m}_{\rm loc}(\tau)$ is the local one-particle Green's function for orbital $m$
and imaginary times $\tau$ at inverse temperature $\beta=1/T$.
The Ru-283 sites are much stronger correlated than e.g. the Ru-99 or Ru-114 sites with
small $\Delta$. In fact, for $U=3.6$\,eV the former sites have already gapped Ru-$t_{2g}$
states, while on average, the Ru sublattice still shows metallic response (see inset in 
top of Fig.~\ref{fig:ex}). The wide spectrum of CF values based on the significant spatial 
inhomogeneity introduced by Ti doping therefore gives rise to a site-selective Mott 
scenario~\cite{par12,lec15} in Ca$_3$(Ru$_{0.9375}$Ti$_{0.0625}$)$_2$O$_7$. 
It occurs on the Ru sites perpendicular-adjacent to the Ti impurities and is 
precursory to the MIT of the complete system at $U\sim 4$\,eV.

Nonsurprisingly, the orbital polarization, here simply defined as
$p=n_{xy}-(n_{xz}+n_{yz})/2$, is much weaker for small-$\Delta$ sites. Those sites
also cause site-selective physics, namely an oscillatory-in-$U$ revival of 
metallicity for $U>4$\,eV. Only for $U=6$\,eV all sites behave insulating 'in line'. 
The given intermediate large $U$-regime can be traced back to strong orbital(-order) 
competition originating from the small-$\Delta$ Ru sites. On average close to the 
metal-insulator transition, the $xy$ orbital becomes completely filled with two electrons, 
whereas the $xz,yz$ become half filled, resulting in $p=1$. However, e.g. the Ru-99 sites 
mark a dominant $xz$ and close to half-filled $yz,xy$ orbital filling for $U=4.75$\,eV 
(see bottom of Fig.~\ref{fig:ex}).
\begin{figure}[t]
\includegraphics*[width=8.5cm]{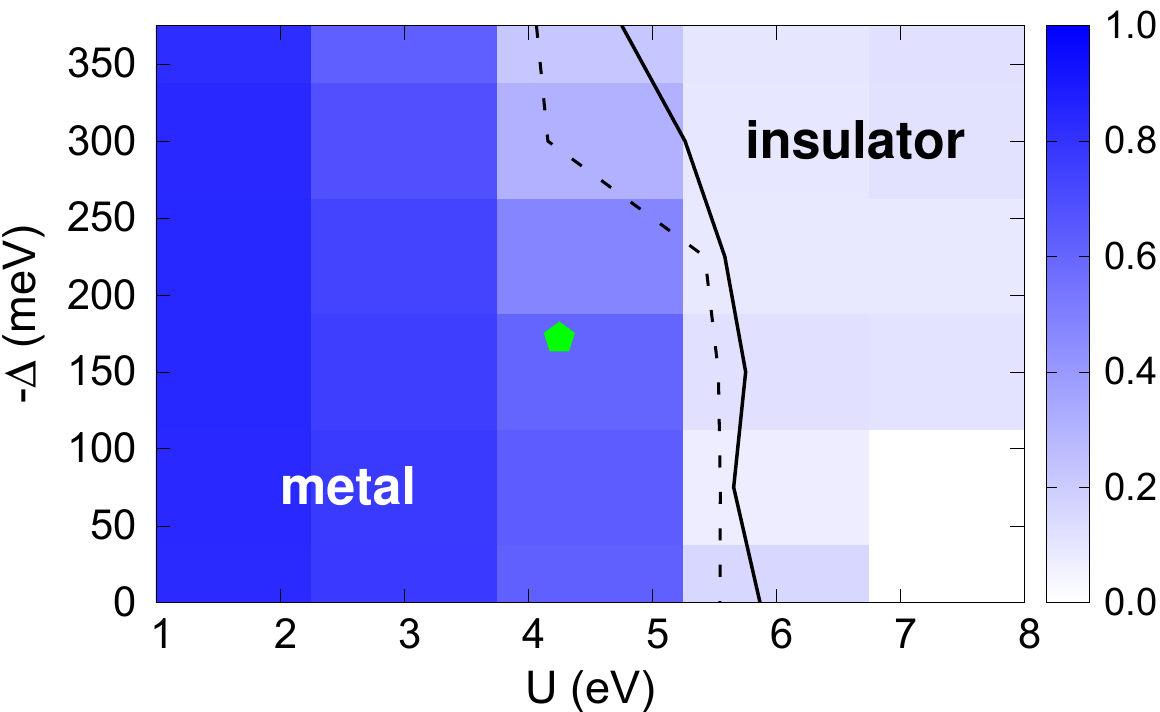}
\caption{Crystal-field vs. $U$ phase diagram for Ti-doped Ca$_3$Ru$_2$O$_7$
based on the EA structure at $T=100$\,K (see text for further details). Blue color scale 
amounts to $A_0$ of Ru$(4d)$-$t_{2g}$ in the PM regime. Full(Dashed) line marks the 
interpolated metal-insulator phase boundary, here roughly defined by $A_0=0.2$, 
in the PM(G-AFM) regime. Green pentagon marks the position of the onset of the insulating 
phase in the supercell calculations (here located via its average $\Delta$ value).}
\label{fig:dia}
\end{figure}

In order to render the comparison between the spatial homogeneous and -inhomogeneous case
more explicit, Fig.~\ref{fig:dia} displays the $-\Delta$ vs. $U$ phase diagram of 
effective Ti-doped Ca$_3$Ru$_2$O$_7$. It is constructed by utilizing the EA structure and 
gradually fixing $\Delta$ within the DFT+DMFT calculations. Note that charge 
self-consistency is an important ingredient, since orbital polarization and MIT 
tendencies depend thereon~\cite{lec18}. Since the true experimental system is AFM ordered 
in the insulating phase, results for G-AFM order of the EA structure are included.
Those were obtained by enabling a spin-polarized self-energy in the DFT+DMFT cycle. 
As expected, the MIT occurs at somewhat smaller $U$ with G-AFM order, yet
the net effect of magnetic ordering on the Mott criticality is not decisive.
Increasing the absolute value of $\Delta$ fosters the driving toward the MIT, but the slope 
remains steep for $-\Delta<200$\,meV. The MIT occurs for moderate $U$ values only for 
rather large CF splittings. Note in this respect that a paramagnetic MIT was 
realized for single-layer Ca$_2$RuO$_4$ with $\Delta\sim -320$\,meV for $U=3.1$\,eV 
in a previous one-shot DFT+DMFT calculation~\cite{gor10}. But such one-shot approaches
are known for two features compared to charge self-consistent DFT+DMFT~\cite{lec18,ham20}, 
namely tending to overestimate orbital polarizations and often realizing smaller critical $U$ 
values. Furthermore, the MIT value of $U=4.25$\,eV for the defect supercell (see green 
pentagon in Fig.~\ref{fig:dia}) is still well in the metallic region of the EA-based phase 
diagram. Yet remember that the 'orbital-competing regime' up to a robust spatially-coherent 
insulating state extends also up to $U=6$\,eV in the supercell calculations. Still, it 
may be inferred that spatial inhomogeneity, which is not included in the EA structure, is 
proactive in driving the MIT toward smaller $U$ values.

We moreover performed DFT+DMFT calculations with allowing for AFM ordering in the
Ti-doped supercell. Also there, competing orderings, now between different AFM
tendencies occur: while or moderate $3<U<4$\,eV the system tends to A-type like ordering, 
reminiscent of the stoichiometric magnetic order, for $U>4$\,eV indeed G-type like 
ordering becomes more favorable. Above $U=6$\,eV a very robust G-AFM ordering pattern
is established. 

{\it Summary.---}
By means of large-scale first-principles many-body calculations, we showed that the
MIT in Ti-doped Ca$_3$Ru$_2$O$_7$ is mainly driven by the interplay of spatial-inhomogeneity
features and strong electronic correlations. Introduction of Ti impurities leads to a
substantial energy spread of the CF splitting $\Delta$ across the Ru sublattice, resulting in 
site-selective Mott transitions for large-$\Delta$ sites and competitions in the 
orbital-ordering tendencies on small-$\Delta$ sites. This explains not only the occurrence
of a doping-induced metal-insulator transition for reasonable interaction strengths, but
accounts furthermore for the findings of perculative behavior. Other effects such as 
hopping-blocking or average structural modification via a change of lattice parameters
furthermore support the insulating tendencies with doping, but appear not decisive. 
The here reported site-selective physics may also be relevant for an understanding and
modelling of recently discovered current-induced diamagnetism~\cite{sow19} in 
Ti-doped Ca$_3$Ru$_2$O$_7$.
Describing the transport scenario from a possibly current-induced (partial-)gap closing 
and/or the coexistence of diamagnetism with persisting (local) magnetic order, could 
rely on the important interplay of electronic correlation and spatial inhomogeneity.
Furthermore concerning experiments, scanning-tunneling studies of 
Ca$_3$(Ru$_{1-x}$Ti$_x$)$_2$O$_7$ would be highly interesting to verify the here unveiled
site-dependent Mott physics.

Let us finally emphasize that sole averaged-structure investigations of defect problems 
in correlated materials may have their shortcomings. In a recent realistic DMFT study on 
impurities in V$_2$O$_3$~\cite{lec18}, it was shown that local point-group symmetry
breaking from trigonal to monoclinic is essential to understand the Cr-induced paramagnetic
MIT. And here, Ti-induced site-symmetry breakings are key to the MIT in
Ca$_3$Ru$_2$O$_7$. Both results corroborate that the explicit and detailed cooperation 
of defect chemistry and many-body physics is at the heart of various doping problems in 
correlated matter.

\begin{acknowledgements}
F.L. acknowledges financial support from the German Science Foundation (DFG) via the 
project LE-2446/4-1. A.J.M. acknowledges support from the Basic Energy Sciences program 
of the US Department of Energy under grant SC-0018218.
Computations were performed at the University of Hamburg and the JUWELS 
Cluster of the J\"ulich Supercomputing Centre (JSC) under project number hhh08.
\end{acknowledgements}

\bibliographystyle{apsrev4-1}
\bibliography{327ref}

\end{document}